\documentclass[conference]{IEEEtran}
\usepackage{fancyhdr, epsfig, epsf, amsthm, amsmath, amssymb, amsfonts, subfigure, color}
\usepackage{threeparttable}
\usepackage[noadjust]{cite}
\usepackage{dsfont}
\usepackage{enumerate}
\usepackage{comment}
\usepackage{soul}
\usepackage{kotex}
\usepackage{algorithm}
\usepackage{algpseudocode}
\usepackage[thinc]{esdiff}
\usepackage{multirow}
\usepackage{makecell}

\makeatletter
\newcommand{\algmargin}{\the\ALG@thistlm}
\makeatother
\algnewcommand{\parState}[1]{\State%
    \parbox[t]{\dimexpr\linewidth-\algmargin}{\hangindent=\algorithmicindent \hangafter=1 #1\strut}}

\usepackage{cite}
\usepackage{graphicx}
\usepackage{textcomp}
\usepackage{xcolor}
\usepackage[scr = zapfc]{mathalfa}

\includecomment{comment}	

\newcolumntype{C}[1]{>{\centering\arraybackslash}p{#1}}

\def\BibTeX{{\rm B\kern-.05em{\sc i\kern-.025em b}\kern-.08em
    T\kern-.1667em\lower.7ex\hbox{E}\kern-.125emX}}

\usepackage{mathtools}
\usepackage[noabbrev]{cleveref}

\IEEEoverridecommandlockouts

\begin{document}
\newtheorem{theorem}{Theorem}
\newtheorem{acknowledgement}[theorem]{Acknowledgement}
\newtheorem{axiom}[theorem]{Axiom}
\newtheorem{case}[theorem]{Case}
\newtheorem{claim}[theorem]{Claim}
\newtheorem{conclusion}[theorem]{Conclusion}
\newtheorem{condition}[theorem]{Condition}
\newtheorem{conjecture}[theorem]{Conjecture}
\newtheorem{criterion}[theorem]{Criterion}
\newtheorem{definition}{Definition}
\newtheorem{exercise}[theorem]{Exercise}
\newtheorem{lemma}{Lemma}
\newtheorem{corollary}{Corollary}
\newtheorem{notation}[theorem]{Notation}
\newtheorem{problem}[theorem]{Problem}
\newtheorem{proposition}{Proposition}
\newtheorem{remark}{Remark}
\newtheorem{solution}[theorem]{Solution}
\newtheorem{summary}[theorem]{Summary}
\newtheorem{assumption}{Assumption}
\newtheorem{example}{\bf Example}
\newtheorem{probform}{\bf Problem}

\def\qed{$\Box$}
\def\QED{\mbox{\phantom{m}}\nolinebreak\hfill$\,\Box$}
\def\proof{\noindent{\emph{Proof:} }}
\def\poof{\noindent{\emph{Sketch of Proof:} }}
\def
\endproof{\hspace*{\fill}~\qed
\par
\endtrivlist\unskip}
\def\endproof{\hspace*{\fill}~\qed\par\endtrivlist\vskip3pt}

\def\E{\mathsf{E}}
\def\eps{\varepsilon}
\def\Lsp{{\boldsymbol L}}
\def\Bsp{{\boldsymbol B}}
\def\lsp{{\boldsymbol\ell}}
\def\Ltsp{{\Lsp^2}}
\def\Lpsp{{\Lsp^p}}
\def\Linsp{{\Lsp^{\infty}}}
\def\LtR{{\Lsp^2(\Rst)}}
\def\ltZ{{\lsp^2(\Zst)}}
\def\ltsp{{\lsp^2}}
\def\ltZt{{\lsp^2(\Zst^{2})}}
\def\ninN{{n{\in}\Nst}}
\def\oh{{\frac{1}{2}}}
\def\grass{{\cal G}}
\def\ord{{\cal O}}
\def\dist{{d_G}}
\def\conj#1{{\overline#1}}
\def\ntoinf{{n \rightarrow \infty }}
\def\toinf{{\rightarrow \infty }}
\def\tozero{{\rightarrow 0 }}
\def\trace{{\operatorname{trace}}}
\def\ord{{\cal O}}
\def\UU{{\cal U}}
\def\rank{{\operatorname{rank}}}
\def\acos{{\operatorname{acos}}}

\def\SINR{\mathsf{SINR}}
\def\SNR{\mathsf{SNR}}
\def\SIR{\mathsf{SIR}}
\def\tSIR{\widetilde{\mathsf{SIR}}}
\def\Ei{\mathsf{Ei}}
\def\l{\left}
\def\r{\right}
\def\({\left(}
\def\){\right)}
\def\lb{\left\{}
\def\rb{\right\}}

\setcounter{page}{1}

\newcommand{\eref}[1]{(\ref{#1})}
\newcommand{\fig}[1]{Fig.\ \ref{#1}}

\def\bydef{:=}
\def\ba{{\mathbf{a}}}
\def\bb{{\mathbf{b}}}
\def\bc{{\mathbf{c}}}
\def\bd{{\mathbf{d}}}
\def\bee{{\mathbf{e}}}
\def\bff{{\mathbf{f}}}
\def\bg{{\mathbf{g}}}
\def\bh{{\mathbf{h}}}
\def\bi{{\mathbf{i}}}
\def\bj{{\mathbf{j}}}
\def\bk{{\mathbf{k}}}
\def\bl{{\mathbf{l}}}
\def\bm{{\mathbf{m}}}
\def\bn{{\mathbf{n}}}
\def\bo{{\mathbf{o}}}
\def\bp{{\mathbf{p}}}
\def\bq{{\mathbf{q}}}
\def\br{{\mathbf{r}}}
\def\bs{{\mathbf{s}}}
\def\bt{{\mathbf{t}}}
\def\bu{{\mathbf{u}}}
\def\bv{{\mathbf{v}}}
\def\bw{{\mathbf{w}}}
\def\bx{{\mathbf{x}}}
\def\by{{\mathbf{y}}}
\def\bz{{\mathbf{z}}}
\def\b0{{\mathbf{0}}}

\def\bA{{\mathbf{A}}}
\def\bB{{\mathbf{B}}}
\def\bC{{\mathbf{C}}}
\def\bD{{\mathbf{D}}}
\def\bE{{\mathbf{E}}}
\def\bF{{\mathbf{F}}}
\def\bG{{\mathbf{G}}}
\def\bH{{\mathbf{H}}}
\def\bI{{\mathbf{I}}}
\def\bJ{{\mathbf{J}}}
\def\bK{{\mathbf{K}}}
\def\bL{{\mathbf{L}}}
\def\bM{{\mathbf{M}}}
\def\bN{{\mathbf{N}}}
\def\bO{{\mathbf{O}}}
\def\bP{{\mathbf{P}}}
\def\bQ{{\mathbf{Q}}}
\def\bR{{\mathbf{R}}}
\def\bS{{\mathbf{S}}}
\def\bT{{\mathbf{T}}}
\def\bU{{\mathbf{U}}}
\def\bV{{\mathbf{V}}}
\def\bW{{\mathbf{W}}}
\def\bX{{\mathbf{X}}}
\def\bY{{\mathbf{Y}}}
\def\bZ{{\mathbf{Z}}}

\def\mA{{\mathbb{A}}}
\def\mB{{\mathbb{B}}}
\def\mC{{\mathbb{C}}}
\def\mD{{\mathbb{D}}}
\def\mE{{\mathbb{E}}}
\def\mF{{\mathbb{F}}}
\def\mG{{\mathbb{G}}}
\def\mH{{\mathbb{H}}}
\def\mI{{\mathbb{I}}}
\def\mJ{{\mathbb{J}}}
\def\mK{{\mathbb{K}}}
\def\mL{{\mathbb{L}}}
\def\mM{{\mathbb{M}}}
\def\mN{{\mathbb{N}}}
\def\mO{{\mathbb{O}}}
\def\mP{{\mathbb{P}}}
\def\mQ{{\mathbb{Q}}}
\def\mR{{\mathbb{R}}}
\def\mS{{\mathbb{S}}}
\def\mT{{\mathbb{T}}}
\def\mU{{\mathbb{U}}}
\def\mV{{\mathbb{V}}}
\def\mW{{\mathbb{W}}}
\def\mX{{\mathbb{X}}}
\def\mY{{\mathbb{Y}}}
\def\mZ{{\mathbb{Z}}}

\def\cA{\mathcal{A}}
\def\cB{\mathcal{B}}
\def\cC{\mathcal{C}}
\def\cD{\mathcal{D}}
\def\cE{\mathcal{E}}
\def\cF{\mathcal{F}}
\def\cG{\mathcal{G}}
\def\cH{\mathcal{H}}
\def\cI{\mathcal{I}}
\def\cJ{\mathcal{J}}
\def\cK{\mathcal{K}}
\def\cL{\mathcal{L}}
\def\cM{\mathcal{M}}
\def\cN{\mathcal{N}}
\def\cO{\mathcal{O}}
\def\cP{\mathcal{P}}
\def\cQ{\mathcal{Q}}
\def\cR{\mathcal{R}}
\def\cS{\mathcal{S}}
\def\cT{\mathcal{T}}
\def\cU{\mathcal{U}}
\def\cV{\mathcal{V}}
\def\cW{\mathcal{W}}
\def\cX{\mathcal{X}}
\def\cY{\mathcal{Y}}
\def\cZ{\mathcal{Z}}
\def\cd{\mathcal{d}}
\def\Mt{M_{t}}
\def\Mr{M_{r}}
\def\O{\Omega_{M_{t}}}
\newcommand{\figref}[1]{{Fig.}~\ref{#1}}
\newcommand{\tabref}[1]{{Table}~\ref{#1}}

\newcommand{\var}{\mathsf{var}}
\newcommand{\fb}{\tx{fb}}
\newcommand{\nf}{\tx{nf}}
\newcommand{\BC}{\tx{(bc)}}
\newcommand{\MAC}{\tx{(mac)}}
\newcommand{\Pout}{p_{\mathsf{out}}}
\newcommand{\nnn}{\nn\\}
\newcommand{\FB}{\tx{FB}}
\newcommand{\TX}{\tx{TX}}
\newcommand{\RX}{\tx{RX}}
\renewcommand{\mod}{\tx{mod}}
\newcommand{\m}[1]{\mathbf{#1}}
\newcommand{\td}[1]{\tilde{#1}}
\newcommand{\sbf}[1]{\scriptsize{\textbf{#1}}}
\newcommand{\stxt}[1]{\scriptsize{\textrm{#1}}}
\newcommand{\suml}[2]{\sum\limits_{#1}^{#2}}
\newcommand{\sumlk}{\sum\limits_{k=0}^{K-1}}
\newcommand{\eqhsp}{\hspace{10 pt}}
\newcommand{\tx}[1]{\texttt{#1}}
\newcommand{\Hz}{\ \tx{Hz}}
\newcommand{\sinc}{\tx{sinc}}
\newcommand{\tr}{\mathrm{tr}}
\newcommand{\diag}{\mathrm{diag}}
\newcommand{\MAI}{\tx{MAI}}
\newcommand{\ISI}{\tx{ISI}}
\newcommand{\IBI}{\tx{IBI}}
\newcommand{\CN}{\tx{CN}}
\newcommand{\CP}{\tx{CP}}
\newcommand{\ZP}{\tx{ZP}}
\newcommand{\ZF}{\tx{ZF}}
\newcommand{\SP}{\tx{SP}}
\newcommand{\MMSE}{\tx{MMSE}}
\newcommand{\MINF}{\tx{MINF}}
\newcommand{\RC}{\tx{MP}}
\newcommand{\MBER}{\tx{MBER}}
\newcommand{\MSNR}{\tx{MSNR}}
\newcommand{\MCAP}{\tx{MCAP}}
\newcommand{\vol}{\tx{vol}}
\newcommand{\ah}{\hat{g}}
\newcommand{\tg}{\tilde{g}}
\newcommand{\teta}{\tilde{\eta}}
\newcommand{\heta}{\hat{\eta}}
\newcommand{\uh}{\m{\hat{s}}}
\newcommand{\eh}{\m{\hat{\eta}}}
\newcommand{\hv}{\m{h}}
\newcommand{\hh}{\m{\hat{h}}}
\newcommand{\Po}{P_{\mathrm{out}}}
\newcommand{\Poh}{\hat{P}_{\mathrm{out}}}
\newcommand{\Ph}{\hat{\gamma}}
\newcommand{\mat}[1]{\begin{matrix}#1\end{matrix}}
\newcommand{\ud}{^{\dagger}}
\newcommand{\C}{\mathcal{C}}
\newcommand{\nn}{\nonumber}
\newcommand{\nInf}{U\rightarrow \infty}

\setlength{\textfloatsep}{0pt}
\setlength{\belowdisplayskip}{1pt} 

\title{\LARGE Joint Data Deepening-and-Prefetching 
for Energy-Efficient Edge Learning
}


\author{\IEEEauthorblockN{Sujin Kook$^*$, Won-Yong Shin$^\ddagger$, Seong-Lyun Kim$^*$, Seung-Woo Ko$^\S$}\\
\IEEEauthorblockA{$^*$School of EEE, {Yonsei University}, Seoul, Korea, email: \{sjkook, slkim\}@ramo.yonsei.ac.kr}
\IEEEauthorblockA{$^\ddagger$Dept. of Comput. Science and Eng., {Yonsei University}, Seoul, Korea, email: wy.shin@yonsei.ac.kr}
\IEEEauthorblockA{$^\S$Dept. of Smart Mobility Eng., {Inha University}, Incheon, Korea, email: swko@inha.ac.kr}
}

\maketitle

\begin{abstract}
The vision of pervasive \emph{machine learning} (ML) services can be realized by training an ML model on time using real-time data collected by \emph{internet of things} (IoT) devices. To this end, IoT devices require offloading their data to an edge server in proximity. On the other hand, high dimensional data with a heavy volume causes a significant burden to an IoT device with a limited energy budget. To cope with the limitation, we propose a novel offloading architecture, called \emph{joint data deepening and prefetching} (JD2P), which is feature-by-feature offloading comprising two key techniques. The first one is \emph{data deepening}, where each data sample's features are sequentially offloaded in the order of importance determined by the data embedding technique such as \emph{principle component analysis} (PCA). No more features are offloaded when the features offloaded so far are enough to classify the data, resulting in reducing the amount of offloaded data. The second one is \emph{data prefetching}, where some features potentially required in the future are offloaded in advance, thus achieving high efficiency via precise prediction and parameter optimization. To verify the effectiveness of JD2P, we conduct experiments using the MNIST and fashion-MNIST  dataset. Experimental results demonstrate that the JD2P can significantly reduce the expected energy consumption compared with several benchmarks without degrading learning accuracy.
\end{abstract}

\vspace{5pt}
\section{Introduction}
\vspace{5pt}
With the wide spread of \emph{internet of things} (IoT) devices, a huge amount of real-time data have been continuously generated. It can be fuel for operating various on-device \emph{machine learning} (ML) services, e.g., object detection and natural language processing, if provided on time. One viable technology to this end is \emph{edge learning}, where an ML model is trained at the edge server in proximity using the data offloaded from IoT devices \cite{zhu2020toward}. Compared to the learning at the cloud server, IoT devices can offer the latest data to the edge server before out-of-date, and the resultant ML model can reflect the current environment precisely without a dataset shift \cite{quinonero2008dataset} or catastrophic forgetting \cite{goodfellow2013empirical}.

On the other hand, as the concerned environment becomes complex, the data collected by each device tends to be high-dimensional with heavy volume, thus causing a significant burden to offload data for an IoT device with a limited energy budget. Several attempts have been proposed in the literature to address this issue, whose main thrust is to selectively offload data depending on the importance of data to the concerned ML model. In \cite{liu2020wireless}, motivated by the classic \emph{support vector machine} (SVM) technique, data importance was defined inversely proportional to its uncertainty, which corresponds to the margin to the decision boundary. A selective retransmission decision was optimized by allowing more transmissions for data with high uncertainty, leading to the corresponding ML model's fast convergence. In the same vein, the scheduling issue of multi-device edge learning has been tackled in \cite{liu2020data}, where a device having more important data samples is granted access to the medium more frequently. In \cite{he2020importance}, a data sample's gradient norm obtained during training a \emph{deep neural network} (DNN) was regarded as the corresponding importance metric. It enables each mobile device to select data sample that is likely to contribute to its local ML model training in a federated edge learning system. In \cite{taik2021data}, data importance was defined at the dispersed level of dataset distribution. A device with an important dataset is allowed to assign more bandwidth to accelerate the training process.

Aligned with the trend, we aim to develop a novel edge learning architecture, called \emph{joint data deepening and prefetching} (JD2P). The above prior works quantify the importance of each data sample or the entire dataset, bringing about a significant communication overhead when raw data become complex with a higher dimension. On the other hand, the proposed JD2P leverages the technique of data embedding to extract a few features from raw data and sort them in the order of importance. This allows us to design a feature importance-based offloading technique, called data deepening;  Features are sequentially offloaded in the important order and stop offloading the next one if reaching the desired performance. Besides, several data samples' subsequent features can be offloaded proactively before requested, called data prefetching, which extends the offloading duration and thus achieves higher energy efficiency. Through relevant parameter optimizations and extensive simulation studies using the MNIST and fashion-MNIST dataset, it is verified that the JD2P reduces the expected energy consumption significantly than several benchmarks without degrading learning accuracy.   

\begin{figure}[t]
\centering
\centering
\includegraphics[width=8cm]{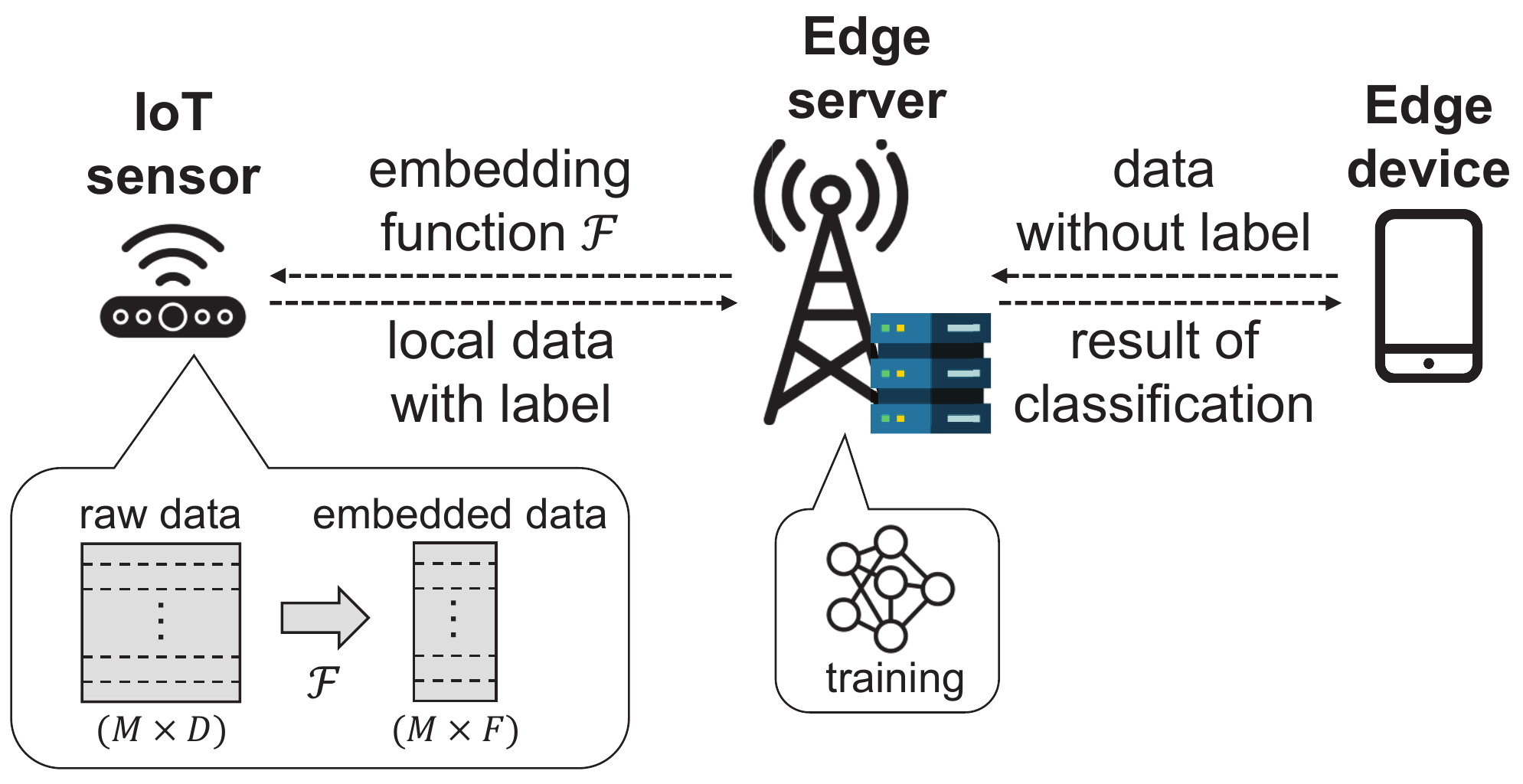}
\vspace{-5pt}
\caption{Edge learning network comprising a pair of edge device and edge server collocated with a wireless access point. }
\label{Fig1-system_model}
\vspace{10pt}
\end{figure}

\section{System Model} \label{section:system model}

This section describes our system model, including the concerned scenario, data structure, and offloading model.

\subsection{Edge Learning Scenario}

Consider an edge learning network comprising a pair of the edge server and the IoT sensor as a data collector (see Fig. \ref{Fig1-system_model}). 
We aim at training a binary classifier using local data with two classes collected by the IoT sensor. Due to the IoT sensor's limited computation capability, the edge server is requested to train a classifier with the local data offloaded from IoT sensor instead of training the classifier on the IoT sensor.

\subsection{Data Embedding}

Consider $M$ samples measured at the sensor, denoted by $\mathbf{y}_m\in\mathbb{R}^D$, where $m$ is the index of data sample, i.e., $m=1,2,\cdots, M$. We assume that the class of each sample is known, denoted by ${c}_m\in\{0,1\}$. Each raw data sample's dimension, say $D$, is assumed to be equivalent. The dimension $D$ is in general sufficiently high to reflect complex environments, which is known as an obstacle to achieve high-accuracy classification \cite{domingos2012few}. Besides, a large amount of energy is required to offload these raw data to the edge server. To overcome these limitations, these high dimensional raw data can be embedded into a low-dimensional space using data embedding techniques  \cite{zheng2014coupled}, such as \emph{principle component analysis} (PCA) \cite{abdi2010principal} and auto-encoder \cite{hinton2006reducing}. Specifically, given $F$ less than $D$, there exists a mapping function $\mathcal{F}:\mathbb{R}^D\rightarrow \mathbb{R}^F$ such that 
\begin{align}\label{Eq:Data_Embedding}
    \mathbf{x}_m=\mathcal{F}(\mathbf{y}_m),
\end{align}
where $\mathbf{x}_m= \left[x_{m,1}, \cdots , x_{m,F}\right]^T$ represents the embedded data with $F$ features. We assume that the edge device knows the embedding function $\mathcal{F}$, which has been trained by the edge server using the historical data set. We use PCA as a primary feature embedding technique due to its low computational overhead, while other techniques are straightforwardly applicable. Partial or all features of each embedded data are offloaded depending on the offloading and learning designs introduced in the sequel.

\subsection{Offloading Model}\label{subsection:offloading_model}

The entire offloading duration is slotted into $K$ rounds with $t_0$ seconds. The channel gain in round $k$ is denoted as $g_k$ with $g_k>0$. We assume that  channel gains are constant over one time slot and \emph{independently and identically distributed} (i.i.d.) over different rounds. Following the models in \cite{tao2019stochastic} and \cite{E2_zhang2013energy}, the transmission power required to transmit $b$ bits in round $k$, denoted as $e_k$, is modeled by a {monomial function} and is given as $e_k = \lambda \frac{{(b/t)}^{\ell}}{g_k}$ where $\lambda$ is the energy coefficient, $\ell$ represents the monomial order, and $t$ is an allowable transmission duration for $b$ bits. The typical range for a monomial order is $2\leq \ell \leq 5$ because this order depends on the specific modulation and coding scheme. Then, the energy consumption in round $k$, which is the product of $e_k$ and $t$, is given as
\begin{align} \label{EQ : E}
    \mathbf{E}(b, t;g_k) = {e_k}{t} = \lambda \frac{{b}^\ell}{g_k\left(t\right)^{\ell-1}}.
\end{align}
It is shown that energy consumption is proportional to the transmitted data size $b$, and inversely proportional to the transmission time $t$. For energy-efficient edge learning, it is necessary to decrease the amount of transmitted data and increase the transmission time.

\section{Joint Data Deepening-and-Prefetching}
This section aims at describing JD2P as a novel architecture to realize energy-efficient edge learning. The overall architecture is briefly introduced first and the detailed techniques of JD2P are elaborated next. 


\subsection{Overview}
 
The proposed JD2P is a feature-by-feature offloading control for energy-efficient classifier training, built on the following definition. 
\begin{definition}[Data Depth]\label{Assumption:FeatureImportance}\emph{A embedded data sample $\mathbf{x}_m$ is said to have depth $k$ when features from $1$ to $k$, say $\mathbf{x}_m^{(k)}=[x_{m,1},\cdots, x_{m,k}]^T$, are enough to correctly predict its class.}
\end{definition}
By Definition \ref{Assumption:FeatureImportance}, we can offload less amount of data required to train the classifier and the resultant energy consumption can be reduced if depths of all data are known in advance. On the other hand, each data sample's depth can be determined after the concerned classifier is trained. Eventually, it is required to process each data's depth identification and classifier training simultaneously to cope with the above recursive relation, which is technically challenging. To this end, we propose two key techniques summarized below.  

\begin{figure}[t] 
\centering
\centering
\includegraphics[width=7cm]{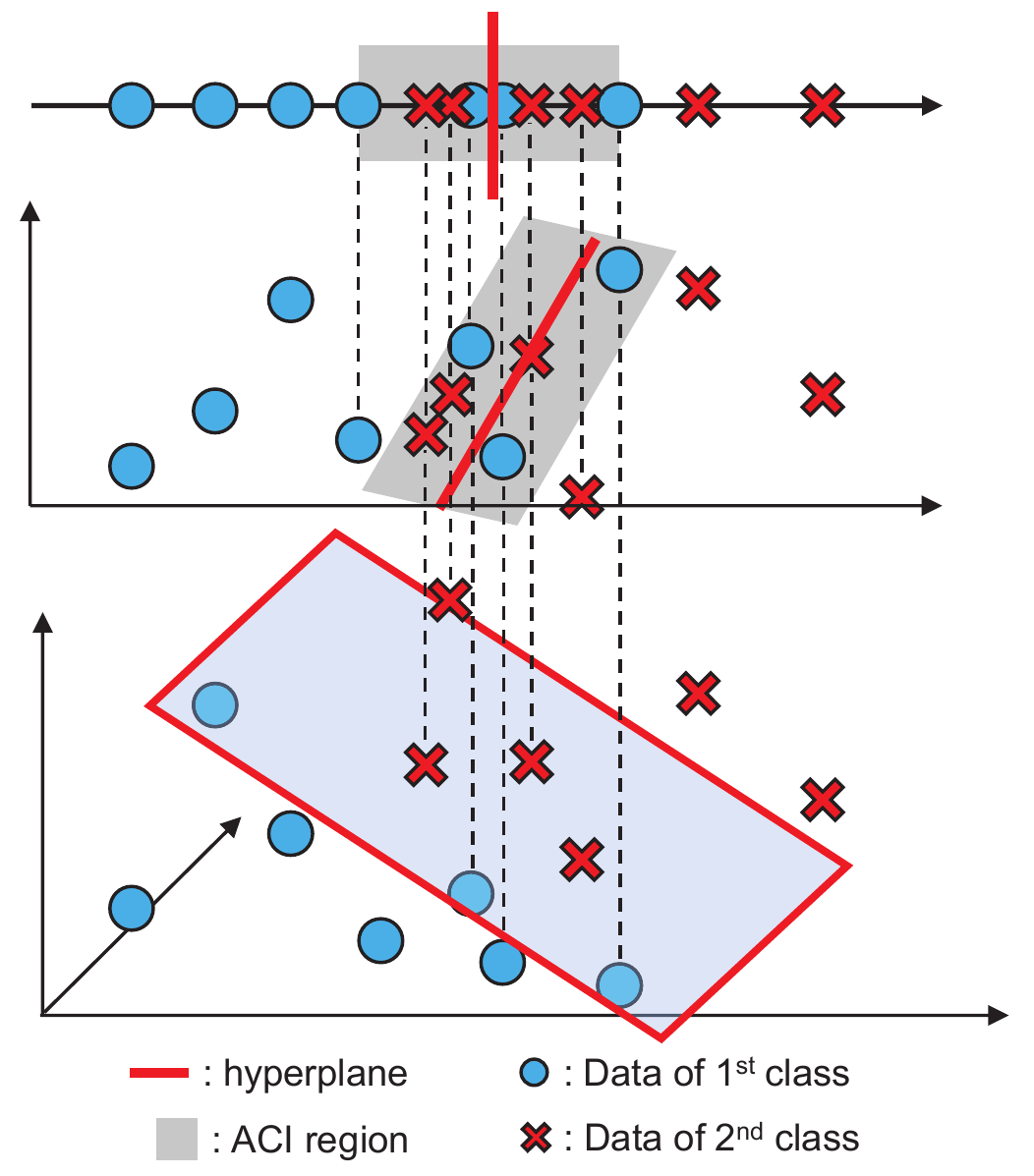}
\caption{The example of data deepening process from 1-dimensional space to 3-dimensional space.}
\label{Fig-Data deepening}
\end{figure}

\subsubsection{Data Deepening} It is a closed-loop offloading decision whether to offload a new feature or not based on the current version of a classifier. Specifically, consider the $k$-depth classifier defined as one trained through features from $1$ to $k$, say $\mathbf{x}_m^{(k)}$ for all $m\in \mathbb{S}^{(k)}$ where $\mathbb{S}^{(k)}$ denotes an index set of data samples that may have a depth of $k$. We use a classic SVM for each depth classifier\footnote{The extension to other classifiers such as DNN and convolutional neural network (CNN) are straightforward, which remains for our future work.}, whose decision hyperplane is given as
\begin{align}\label{Eq: Hyperplane}
    (\mathbf{w}^{(k)})^T\mathbf{x}^{(k)}+b^{(k)} =0, 
\end{align}
where $\mathbf{w}^{(k)} \in \mathbb{R}^k$ is the vector perpendicular to the hyperplane and $b^{(k)}$ is the offset parameter. Given a data sample $\mathbf{x}_m^{(k)}$ for $m \in \mathbb{S}^{(k)}$, the distance to the hyperplane in \eqref{Eq: Hyperplane} can be computed~as
\begin{align}\label{eq:d_m^k}
    d_m^{(k)} = \left| (\mathbf{w}^{(k)})^T\mathbf{x}_m^{(k)} + b^{(k)} \right| / \|\mathbf{w}^{(k)}\|, 
\end{align}
where $\| \cdot \|$ represents the Euclidean norm. The data sample $\mathbf{x}_m$ is said to be a \emph{clearly classified instance} (CCI) by the $k$-depth classifier if $d_m^{(k)}$ is no less than a threshold ${\bar{d}}^{(k)}$ to be specified in Sec. \ref{subsection:threshold_design}. Otherwise, it is said to be a \emph{ambiguous classified instance} (ACI).  In other words, CCIs are depth-$k$ data not requiring an additional feature. Only ACIs are thus included in a new set $\mathbb{S}^{(k+1)}$, given~as
\begin{align} \label{eq : ACI set}
    \mathbb{S}^{(k+1)} = \left\{ m~ |~ d_m^{(k)} \leq \bar{d}^{(k)}, m\in \mathbb{S}^{(k)}\right\}.
\end{align}
As a result, the edge server requests the edge device to offload the next feature $x_{m,k+1}$ for $m\in\mathbb{S}^{(k+1)}$. Fig. \ref{Fig-Data deepening} illustrates the graphical example of data deepening from $1$-dimensional to $3$-dimensional spaces. The detailed process is summarized in Algorithm \ref{algorithm:deepening} except the design of the threshold ${\bar{d}}^{(k)}$.

\begin{algorithm}[t]
\caption{Data Deepening}
\begin{algorithmic}[1]
    \Require Embedded data $\mathbf{x}_m$ for all $m\in\{1,\cdots, M\}$.
    \State Setting $k = 0$, $\mathbb{S}^{(1)}=\{1,\cdots,M\} $.
    \While{$k \leq K$}
    \State $k = k + 1$.
    \parState{%
    Using $\{\mathbf{x}_m^{(k)}\}$ for $m \in \mathbb{S}^{(k)}$, compute the hyperplane of the $k$-depth classifier, specified in \eqref{Eq: Hyperplane}.}
    \State Compute the threshold ${\bar{d}}^{(k)}$ using Algorithm \ref{algorithm:threshold}. 
    \For {$m \in \mathbb{S}^{(k)}$}
        \State Compute $d_m^{(k)}$ using \eqref{eq:d_m^k}.
        \If{$d_m^{(k)} \leq {\bar{d}}^{(k)}$}
            \State $m \in \mathbb{S}^{(k+1)}$.
        \Else
            \State $m \notin \mathbb{S}^{(k+1)}$.
        \EndIf
    \EndFor
    \EndWhile
\end{algorithmic}
\label{algorithm:deepening}
\end{algorithm}

\subsubsection{Data Prefetching} 
\begin{figure}[t]
\centering
\includegraphics[width=9cm]{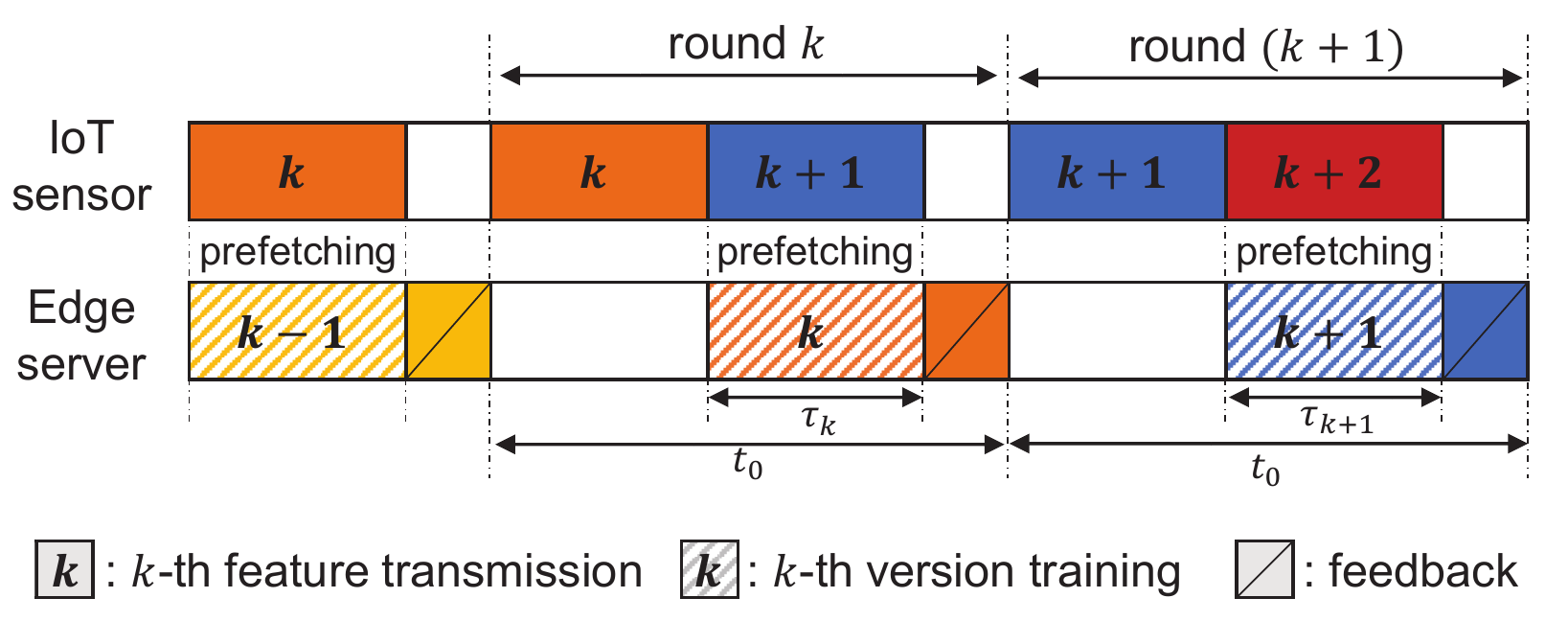}
\vspace{-20pt}
\caption{Data prefetching architecture}  \label{Fig-prefetching}
\end{figure}

As shown in Fig. \ref{Fig-prefetching}, the round $k$ comprises  an offloading duration for the $k$-th features (i.e., $x_{m,k}, \forall m\in\mathbb{S}^{(k)}$) and a training duration for the $k$-depth classifier, and a feedback duration for a new ACI set 
$\mathbb{S}^{(k+1)}$ in \eqref{eq : ACI set}. Without loss of generality, the feedback duration is assumed to be negligible due to its small data size and the edge server's high transmit power. Note that $\mathbb{S}^{(k+1)}$ can be available when starting round $(k+1)$, and a sufficient amount of time should be reserved for training the $(k+1)$-depth classifier. Denote $\tau_{k+1}$ as the corresponding training duration. In other words, the offloading duration $t_{k+1}$ should be no more than $t_0-\tau_{k+1}$, making energy consumption significant as $\tau_{k+1}$ becomes longer.

It can be overcome by offloading the partial data samples' $(k+1)$ features in advance during the training process, called  \emph{prefetching}. The resultant offloading duration can be extended from $t_{k+1}$ to $t_0$, enabling the IoT device to reduce energy consumption, according to \eqref{EQ : E}. On the other hand, the prefetching decision is based on predicting on whether the concerned data sample becomes ACIs. Unless correct, the excessive energy is consumed to prefetch useless features. Balancing the tradeoff is a key, which will be addressed by formulating a stochastic optimization in Sec. \ref{subsection:problem_formulation}.

\subsection{Threshold Design for Data Deepening}
\label{subsection:threshold_design}

\begin{figure}[t] 
\centering
\centering
\includegraphics[width=7.5cm]{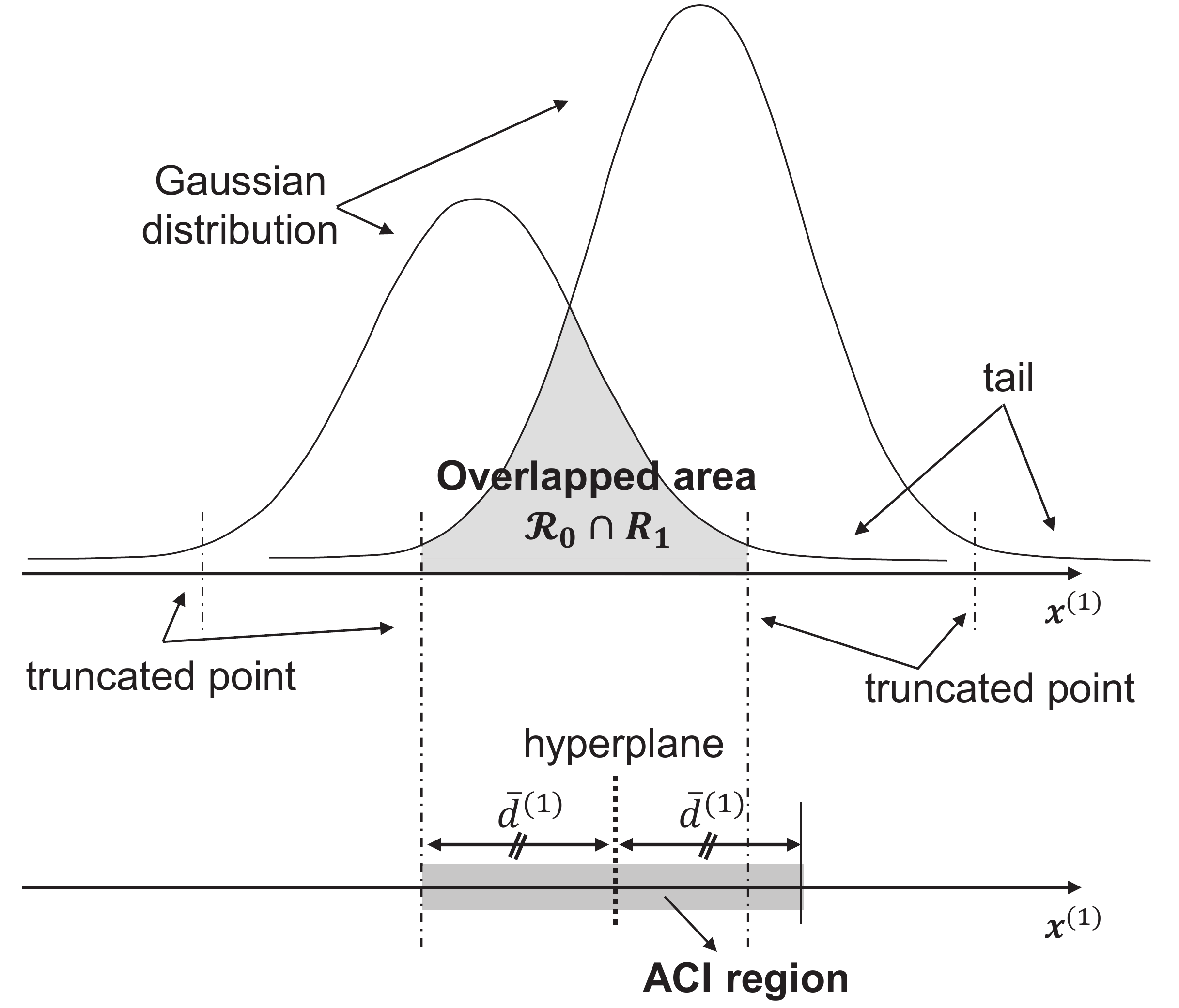}
\vspace{-10pt}
\caption{The ACI region in the 1-dimensional space is obtained by the probability distribution and distance from the hyperplane.}
\label{fig-gaussian}
\end{figure}

This subsection deals with the threshold design 
$\bar{d}^{(k)}$ to categorize whether the concerned data sample $\mathbf{x}_m^{(k)}$ is ACI or CCI based on the $k$-depth classifier.  
The stochastic distribution of each class can be approximated in a form of $k$-variate Gaussian processes using the \emph{Gaussian mixture model} (GMM) \cite{bishop2006pattern}. As shown in Fig. \ref{fig-gaussian}, the overlapped area between two distributions is observed. The data samples in the area are likely to be misclassified. We aim at setting the threshold $\bar{d}^{(k)}$ in such a way that most data samples in the overlapped area are included except a few outliners located in each tail.  

To this end, we introduce the \emph{Mahalanobis distance} (MD) \cite{bensimhoun2009n} as a metric representing the distance from each instant to the concerned distribution. Given class $c\in\{0,1\}$, the MD is defined as
\begin{align}
    \delta_c^{(k)} = \sqrt{\left(\mathbf{x}^{(k)}-\boldsymbol{\mu}_c^{(k)}\right)^T\cdot \left(\boldsymbol{\Sigma}_c^{(k)}\right) ^{-1} \cdot \left(\mathbf{x}^{(k)}-\boldsymbol{\mu}_c^{(k)}\right)}, 
\end{align}
where $\mathbf{x}^{(k)}\sim \mathcal{N}(\boldsymbol{\mu}_c^{(k)}, \boldsymbol{\Sigma}_c^{(k)})$ with $\boldsymbol{\mu}_c^{(k)} \in \mathbb{R}^k$ and  $\mathbf{\Sigma}_c^{(k)} \in \mathbb{R}^{k\times k}$ being the distribution's mean vector and covariance matrix respectively, which are obtainable through the GMM process.
It is obvious that $\delta_c^{(k)}$ is a scale-free random variable and we attempt to set the threshold as the value whose \emph{cumulative distribution function} (CDF) of $\delta_c^{(k)}$ becomes $p_\mathrm{th}$, namely,
\begin{align}
\mathsf{Pr}\left[\delta_c^{(k)}\leq \bar{\delta}_c^{(k)}\right]=p_\mathrm{th}.
\end{align}
Noting that the square of $\delta_c^{(k)}$ follows a chi-square distribution with $k$ degree-of-freedom, the CDF of this distribution for $r>0$ is defined as :
\begin{align}
    \mathscr{G}(r;k) = \mathsf{Pr} \left[ \delta_c^{(k)} \leq r\right] = \frac{\gamma\left(\frac{k}{2},\frac{r}{2}\right)}{\Gamma\left(\frac{k}{2}\right)},
\end{align}
where $\Gamma$ is gamma function defined as $\Gamma(k) = \int_{0}^{\infty} t^{k-1}e^{-t}dt$ and $\gamma$ is the lower incomplete gamma function defined as $\gamma(k, r) = \int_{0}^{r} t^{k-1}e^{-t}dt$. In a closed-form, the threshold $\bar{\delta}_c^{(k)}$ can be given as 
\begin{align}\label{eq:delta_k}
    \bar{\delta}_c^{(k)} = \sqrt{\mathscr{G}^{-1}(p_\mathrm{th};k)},
\end{align}
where $\mathscr{G}^{-1}$ represents the inverse CDF of chi-square distribution with $k$ degree-of-freedom. Due to the scale-free property, the threshold $\bar{\delta}_c^{(k)}$ is identically set regardless of the concerned class; thus, the index of class can be omitted, namely,  $\bar{\delta}_0^{(k)}=\bar{\delta}_1^{(k)}=\bar{\delta}^{(k)}$. Given $\bar{\delta}^{(k)}$, the each distribution can be truncated as 
\begin{align} \label{eq:R_c}
\mathcal{R}_c = \left\{ \mathbf{x}^{(k)}\in\mathbb{R}^{(k)} ~ | ~ \delta_c^{(k)} \leq \bar{\delta}^{(k)}\right\}, \quad c\in\{0,1\}.
\end{align}
Last, the threshold $\bar{d}^{(k)}$ is set by the maximum distance from the hyperplane in \eqref{Eq: Hyperplane} to an arbitrary $k$-dimensional point $\mathbf{x}^{(k)}$ in the overlapped area of $\mathcal{R}_0$ and $\mathcal{R}_1$, given as 
\begin{align} \label{eq:d_k}
    \bar{d}^{(k)} = \max_{\mathbf{x}^{(k)} \in \mathcal{R}_0 \cap \mathcal{R}_1} \left| (\mathbf{w}^{(k)})^T\mathbf{x}^{(k)} + b^{(k)} \right| / \|\mathbf{w}^{(k)}\|.
\end{align}
 The process to obtain the threshold $\bar{d}^{(k)}$ is summarized in Algorithm \ref{algorithm:threshold}.

\begin{remark}[Symmetric ACI Region]\emph{Noting that each class's covariance matrix $\{\mathbf{\Sigma}_c^{(k)}\}_{c\in\{0,1\}}$ is different, the resultant truncated areas of $\mathcal{R}_0$ and $\mathcal{R}_1$ become asymmetric. To avoid the classifier overfitted to one class, we choose the common distance threshold for both classes, say $\bar{d}^{(k)}$ in \eqref{eq:d_k}, corresponding to the maximum distance between the two.}
\end{remark}

\subsection{Hierarchical Edge Inference}
After $K$ rounds, the entire classifier has a hierarchical structure comprising from $1$-depth to $K$-depth classifiers. Consider that a mobile device sends an unlabeled data sample to the edge server, which is initially set as an ACI. Starting from the $1$-depth classifier, the data sample passes through different depth classifiers in sequence until it is changed to an CCI. The last classifier's depth is referred to as the data sample's depth. In other words, its classification result becomes the final one.

\begin{algorithm}[t]
\caption{Finding the threshold $\bar{d}^{(k)}$}
\begin{algorithmic}[1]
    \Require Embedded data $\mathbf{x}_m^{(k)}$ for $m\in \mathbb{S}^{(k)}$, $k$-th version classifier.
    \For{$c \in \{0,1\}$}
        \State Find $\boldsymbol{\mu}_c$, $\boldsymbol{\Sigma}_c$ through GMM process.
        \State Compute $\bar{\delta}_c^{(k)}$ specified in \eqref{eq:delta_k}.
        \parState{%
        Compute the truncated domain of distribution $\mathcal{R}_c$ using \eqref{eq:R_c}}
    \EndFor
    \State Find the overlapped area $\mathcal{R} = \mathcal{R}_0 \bigcap \mathcal{R}_1$.
    \parState{%
    Using $k$-th version classifier in \eqref{Eq: Hyperplane}, compute {$\bar{d}^{(k)}$} using \eqref{eq:d_k}.}
    \\ \Return{$\bar{d}^{(k)}$}.\;
\end{algorithmic}
\label{algorithm:threshold}
\end{algorithm}

\section{Optimal Data Prefetching}\label{section:optimal_data_prefetching}
This section deals with selecting the size of prefetched data in the sense of  minimizing the expected energy consumption of the sensor.

\subsection{Problem Formulation}
\label{subsection:problem_formulation}

Consider the prefetching duration in round $k$, say $\tau_k$, which is equivalent to the training duration of the $k$-depth classifier, as shown in Fig. \ref{Fig-prefetching}. The number of data samples in $\mathbb{S}^{(k)}$ is denoted by $s_k$. Among them, $p_k$ data samples are randomly selected and their $(k+1)$-th features are prefetched.  The prefetched data size is $\alpha p_k$, where $\alpha$ represents the number of bits required to quantize feature data\footnote{The quantization bit rate depends on the value of intensity. For example, one pixel of MNIST data has $255$ intensities and can be quantized 8 bits, i.e., $\alpha = 8$.}. Given the channel gain $g_{k}$, the resultant energy consumption for prefetching is 
\begin{align} \label{EQ :energy(p_k)}
    \mathbf{E}(\alpha p_k, \tau_k ; g_k) = 
    \lambda \frac{\alpha^\ell }{g_{k}\tau_k^{\ell-1}}p_k^\ell.
\end{align}
Here, the number of prefetched data $p_k$ is a discrete control parameter ranging from $0$ to $ s_k$. For tractable optimization in the sequel, we regard $p_k$ as a continuous variable within the range, which is rounded to the nearest integer in practice.   

Next, consider the offloading duration in round $(k+1)$, say $t_{k+1}=t_0-\tau_{k+1}$. Among the data samples in $\mathbb{S}^{(k+1)}$, a few number of data, denoted by $n_{k+1}$, remain after the prefetching. Given the channel gain $g_{k+1}$, the resultant energy consumption is 
\begin{align}
    \mathbf{E}(\alpha n_{k+1}, t_{k+1}; g_{k+1}) = \lambda \frac{\alpha^\ell }{g_{k+1}t_{k+1}^{\ell-1}}n_{k+1}^\ell.
\end{align}
Note that $n_{k+1}$ is determined after the $k$-depth classifier is trained. In other words, $n_{k+1}$ is random at the instant of the prefetching decision. Denote $\rho_k$ as the ratio of a data sample in $\mathbb{S}^{(k)}$ being included in $\mathbb{S}^{(k+1)}$. Then, $n_{k+1}$ follows a binomial distribution with parameters $(s_k-p_k)$ and $\rho_k$, whose probability mass function is $P(j)=\bigl({{s_k-p_k}\atop j}\bigr)\rho_k^j (1-\rho_k)^{s_k-p_k-j}$ for $j=0,\cdots, {s_k-p_k}$. Given $p_k$, the expected energy consumption~is 
\begin{align}
&\mathbb{E}_{n_{k+1}, g_{k+1}}[\mathbf{E}(\alpha n_{k+1}, t_{k+1}; g_{k+1})]
=\lambda\frac{\nu\alpha^\ell}{t_{k+1}^{\ell-1}}\mathbb{E}_{n_{k+1}}[n_{k+1}^\ell],\label{EQ : energy(n_k)}
\end{align}
where $\nu=\mathbb{E}[\frac{1}{g_{k+1}}]$ is the expectation of the inverse channel gain, which can be known a priori due to its i.i.d. property.   

Last, summing up \eqref{EQ :energy(p_k)} and \eqref{EQ : energy(n_k)} is the expected energy consumption for the $(k+1)$-th feature when $p_k$ data samples are prefetched, leading to the following two-stage stochastic optimization: 
\begin{align}
    \min_{p_k} ~ & \lambda \frac{\alpha^\ell }{g_{k}\tau_k^{\ell-1}}p_k^\ell + \lambda\frac{\nu\alpha^\ell}{t_{k+1}^{\ell-1}}\mathbb{E}_{n_{k+1}}[n_{k+1}^\ell]  \nonumber\\
    \text{s.t.} \quad & 0 \leq p_k \leq s_k. \nonumber
    \label{problem:overall} \tag{P1} \nonumber
\end{align}
The optimal prefetching policy can be designed by solving  \ref{problem:overall} explained in the following subsection.

\subsection{Optimal Prefetching Control}
This subsection aims at deriving the closed-form expression of the optimal prefetching number $p_k^*$ by solving \ref{problem:overall}. The main difficulty lies in addressing the $\ell$-th moment $\mathbb{E}_{n_{k+1}}[n_{k+1}^\ell]$, of which the simple form is unknown for general $\ell$. To address it, we refer to the upper bound of the $\ell$-th moment in \cite{2ahle2022sharp},
\begin{align}
    \mathbb{E}_{k+1}[n_{k+1}^{\ell}] \leq \left(\mu_{n_{k+1}} + \frac{\ell}{2}\right)^\ell,
\end{align}
where $\mu_{n_{k+1}} = \left(s_k-p_k\right)\rho_k$ is the mean of the binomial distribution with parameters $(s_k-p_k)$ and $\rho_k$. It is proved in \cite{2ahle2022sharp} that the above upper bound is tight when the order $\ell$ is less than the mean $\mu_{n_{k+1}}$. Therefore, instead of solving \ref{problem:overall} directly, the problem of minimizing the upper bound of the objective function can be formulated as 
\begin{align}
    \min_{p_k} ~& \frac{p_k^\ell}{g_{k}\tau_k^{\ell-1}} + \frac{\nu}{t_{k+1}^{\ell-1}} \left( \left(s_k-p_k\right)\rho_k + \frac{\ell}{2}\right)^\ell
    \nonumber\\
    \text{s.t.} \quad & 0 \leq p_k \leq s_k. \nonumber \tag{P2}\label{Problem:upper bound}
\end{align}
Note that \ref{Problem:upper bound} is a convex optimization, enabling us to derive the closed-form solution. The main result is shown in the following proposition. 
\begin{proposition}[Optimal Prefetching Policy]\label{proposition 1}\emph{Given the ratio of prefetching $\rho_k$ in round $k$, the optimal prefetching data size $p_k^*$, which is the solution to \ref{Problem:upper bound}, is
\begin{align}\label{eq:Optimal_sol}
    p^*_k = \left(\frac{\varphi\rho_k^{\frac{1}{\ell-1}}}{1+\varphi\rho_k^{\frac{\ell}{\ell-1}}}\right) \left(s_k \rho_k + \frac{\ell}{2}\right),
\end{align}
where $\varphi = \left(g_k\nu \right)^{\frac{1}{\ell-1}}\frac{\tau_k}{t_{k+1}}$.}
\end{proposition}
\begin{IEEEproof}
Define the Lagrangian function for \ref{Problem:upper bound} as
\begin{align}
  L = \frac{p_k^\ell}{g_{k}\tau_k^{\ell-1}} + \frac{\nu}{t_{k+1}^{\ell-1}} \left(\left(s_k-p_k\right)\rho_k + \frac{\ell}{2}\right)^\ell + \eta (p_k - s_k),\nonumber
\end{align}
where $\eta$ is a Lagrangian multipliers. Since \ref{Problem:upper bound} is a convex optimization, the following KKT conditions are necessary and sufficient for optimality:
\begin{subequations}\label{eq:kkt}
    \begin{gather}
        \frac{\ell p_k^{\ell-1}}{g_{k}\tau_k^{\ell-1}} - \frac{\ell \nu \rho_k }{t_{k+1}^{\ell-1}} \left(\left(s_k-p_k\right)\rho_k + \frac{\ell}{2}\right)^{\ell-1} +\eta\geq 0, \label{kkt1} \\ 
        p_k \left(\frac{\ell p_k^{\ell-1}}{g_{k}\tau_k^{\ell-1}} - \frac{\ell \nu \rho_k }{t_{k+1}^{\ell-1}} \left(\left(s_k-p_k\right)\rho_k + \frac{\ell}{2}\right)^{\ell-1} +\eta\right) = 0, \label{kkt2} \\ 
        \eta \left(p_k-s_k\right) = 0. \label{kkt3}
    \end{gather}
\end{subequations}
First, if $\eta$ is positive, then $p_k$ should be equal to $s_k$ due to the slackness condition of \eqref{kkt3}, making the LHS of \eqref{kkt2} strictly positive. In other words, the optimal multiplier $\eta$ is zero to satisfy \eqref{kkt2}. Second, with $p_k=0$, the LHS of condition \eqref{kkt1} is always strictly negative unless $\rho_k=0$. As a result, given $\rho_k>0$, $p_k$ should be strictly positive and satisfy the following equality condition: 
\begin{align} \label{eq:last_eq}
  \frac{\ell p_k^{\ell-1}}{g_{k}\tau_k^{\ell-1}} - \frac{\ell \nu \rho_k }{t_{k+1}^{\ell-1}} \left(\left(s_k-p_k\right)\rho_k + \frac{\ell}{2}\right)^{\ell-1}=0. 
\end{align}
Solving \eqref{eq:last_eq} leads to the optimal solution of \ref{Problem:upper bound}, which completes the proof of this proposition. 
\end{IEEEproof}

\begin{remark}[Effect of Parameters]\emph{Assume that the number of ACIs in slot $k$, say $s_k=|\mathbb{S}^{k}|$, is significantly larger than $\frac{\ell}{2}$. We can approximate \eqref{eq:Optimal_sol} as  $p^*_k \approx \left(\frac{\varphi\rho_k^{\frac{1}{\ell-1}}}{1+\varphi\rho_k^{\frac{\ell}{\ell-1}}}\right) s_k \rho_k$. Noting that the term $s_k\rho_k$ represents the expected number of ACIs in slot $(k+1)$, the parameter $\varphi$ controls the portion of prefetching as follows:   
\begin{itemize}
    \item As the current channel gain $g_k$ becomes larger or the training duration $\tau_k$ increases, the parameter $\varphi$ increases and the optimal solution $p_k^*$ reaches near to the $s_k\rho_k$;   
    \item As $g_k$ becomes smaller and $\tau_k$ decreases, both $\varphi$ and $p_k^*$ converge to zero.  
\end{itemize}}
\end{remark}

\section{Simulation Results}\label{Section:Simulation}
In this section, simulation results are presented to validate the superiority of JD2P over several benchmarks. The parameters are set as follows unless stated otherwise. The entire offloading duration consists of $10$ rounds ($K=10$), each of which is set to $t_0 = 0.1$ (sec). For offloading, the channel follows the Gamma distribution with the shape parameter $\beta>1$ and the probability density function $f_g(x) = \frac{x^{\beta-1}e^{-\beta x}}{(1/\beta)^\beta\Gamma(\beta)}$, where the gamma function $\Gamma(\beta) = \int_0^\infty x^{\beta-1}e^{-x}dx$ and the mean $\mathbb{E}[g_k]=1$. The energy coefficient $\lambda$ is set to $10^{-17}$, according to \cite{tao2019stochastic}. The monomial order of the energy consumption model in \eqref{EQ : E} is set as $\ell =3$. For computing and prefetching, the reserved training duration $\tau_k$ is assumed constant for all $k$ and fixed to $\tau_k = \tau =0.5$ (sec) for $1\leq k\leq K$.

We use the MNIST and fashion MNIST datasets for training and testing. Both datasets include $6\cdot 10^4$ training samples and $10^4$ test samples with $784$ gray-scaled pixels. The number of each dataset's classes is $10$. We conduct experiments with every possible pair of classes, namely, $\binom{10}{2}=45$ pairs. PCA is applied for data embedding. For comparison, we consider two benchmark schemes. The first one is to use data deepening only without data prefetching. The second one is full offloading, where all data samples' $10$ features are offloaded first, and the classifier is trained using them. To be specific, the offloading duration of each round is $t_0$ except the last one reduced as $t_{K} = t_0 - \tau$. 

\begin{figure}[t] 
\centering
\centering
\includegraphics[width=8.2cm]{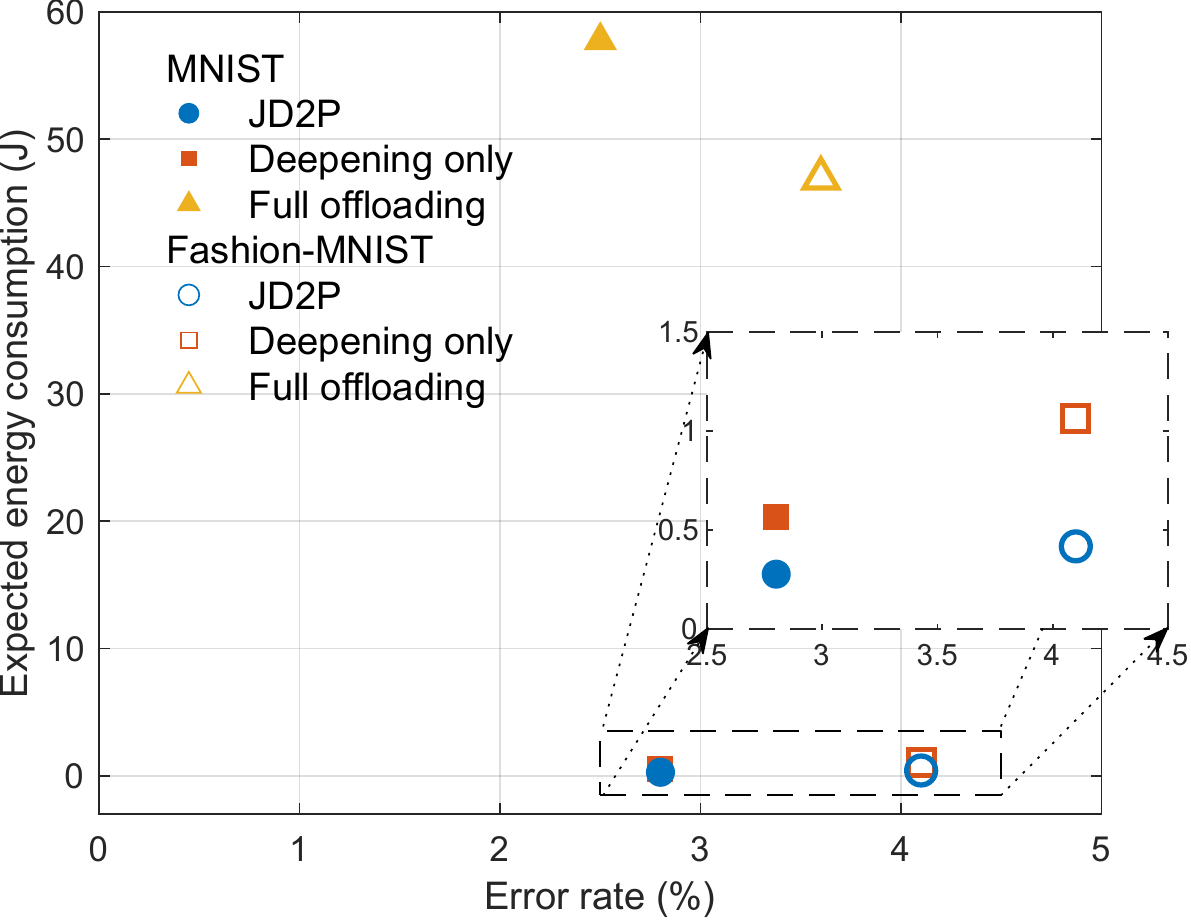}
\vspace{-8pt}
\caption{Performance of JD2P compared with benchmark schemes. Each point represents the average value over all pairs of classes. The training duration is set as $\tau = 0.5$s.}
\label{fig-simulation1}
\vspace{5pt}
\end{figure}

\begin{figure}[t] 
\centering
\centering
\includegraphics[width=7.8cm]{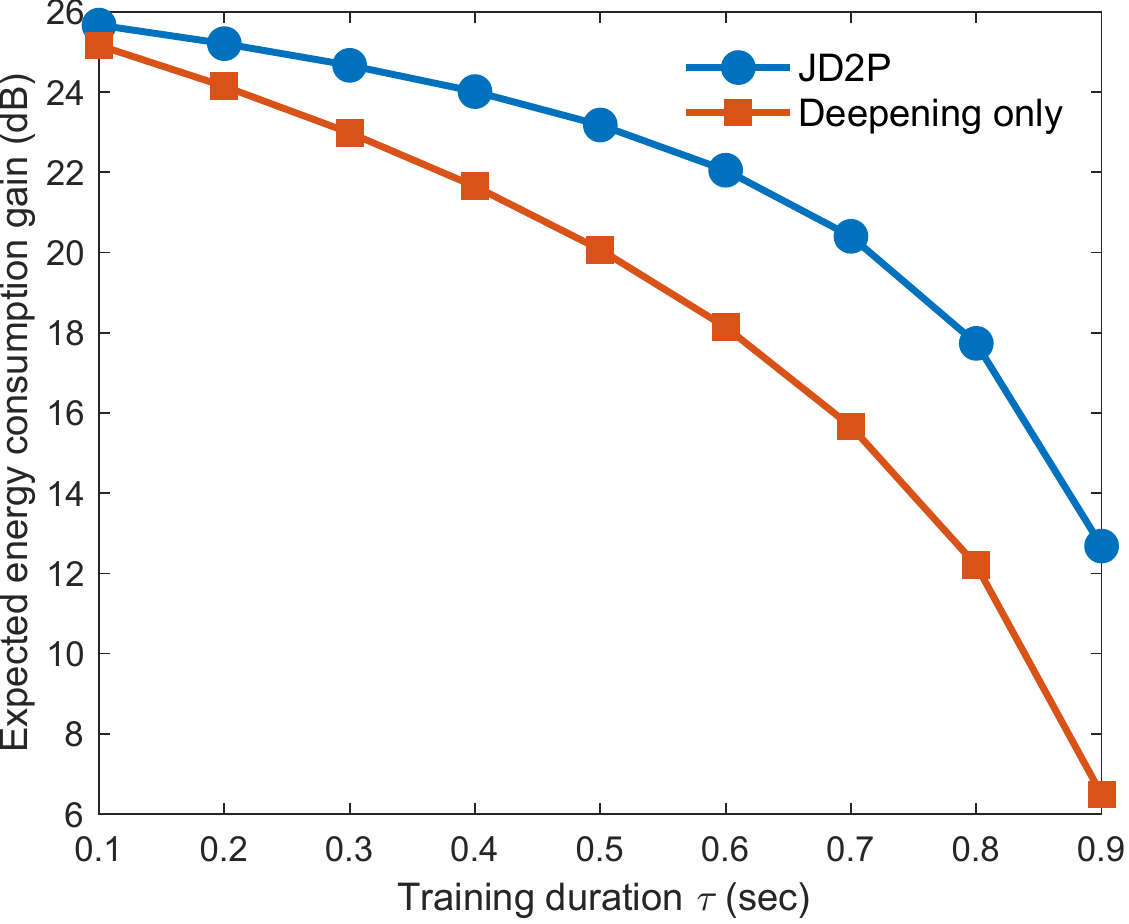}
\vspace{-10pt}
\caption{Effect of the prefetching duration on the expected energy consumption gain obtained by comparing with the full offloading scheme in the case of the  MNIST dataset.}
\label{fig-simulation2}
\end{figure}

First, the expected energy consumption (in Joule) versus the error rate (in \%) is plotted in Fig. \ref{fig-simulation1}. It is shown that the proposed JD2P consumes less energy than the full offloading scheme, namely, $23$dB and $20$dB energy gain for MNIST and fashion-MNIST, respectively, at cost of the marginal degradation in the error rate. The effectiveness of data prefetching is demonstrated in Fig. \ref{fig-simulation2}, plotting the curves of the expected energy consumption gain against the prefetching duration $\tau$ in the case of the MNIST dataset\footnote{The case of the fashion MNIST dataset follows the tendency similar to that of MNIST although the result is omitted in this paper}. The JD2P's expected energy consumption is always smaller than the scheme of data deepening only by sophisticated control of prefetching data in Sec. \ref{section:optimal_data_prefetching}. On the other hand, when compared with the full offloading scheme, the energy gain of JD2P decreases as $\tau$ increases. In other words, a shorter offloading duration compels more data samples to be prefetched, wasting more energy since many prefetched data samples are likely to become CCIs while not being used for the following training.

\section{Concluding remarks}
This study explored the problem of multi-round technique for energy-efficient edge learning. Two criteria for achieving energy efficiency are 1) reducing the amount of offloaded data and 2) extending the offloading duration. JD2P was proposed by addressing both, while integrating data deepening and data prefetching with measuring feature-by-feature data importance and optimizing the amount of prefetched data to avoid wasting energy. Our comprehensive simulation study demonstrated that JD2P can significantly reduce the expected energy consumption compared to several benchmarks.

Though the current work targets to design a simple SVM-based binary classifier with PCA as a key data embedding technique, the proposed JD2P is straightforwardly applicable to more challenging scenarios, such as a multi-class DNN classifier with an advanced data embedding technique. Besides, it is interesting to analyze the performance of JD2P concerning various parameters, which is essential to derive rigorous guidelines for JD2P's practical use.

\end{document}